\newcommand{\ifb}{\ensuremath{\mathrm{fb^{-1}}}}
\newcommand{\TeV}{\ensuremath{\mathrm{Te\kern -0.1em V}}}
\newcommand{\TeVc}{\ensuremath{\mathrm{Te\kern -0.1em V\!/}c}}
\newcommand{\TeVcc}{\ensuremath{\mathrm{Te\kern -0.1em V\!/}c^2}}
\newcommand{\GeV}{\ensuremath{\mathrm{Ge\kern -0.1em V}}}
\newcommand{\GeVc}{\ensuremath{\mathrm{Ge\kern -0.1em V\!/}c}}
\newcommand{\GeVcc}{\ensuremath{\mathrm{Ge\kern -0.1em V\!/}c^2}}
\newcommand{\MeV}{\ensuremath{\mathrm{Me\kern -0.1em V}}}
\newcommand{\MeVc}{\ensuremath{\mathrm{Me\kern -0.1em V\!/}c}}
\newcommand{\MeVcc}{\ensuremath{\mathrm{Me\kern -0.1em V\!/}c^2}}

\newcommand{\um}{\ensuremath{\mathrm{\mu m}}}

\newcommand{\babar}{\mbox{\slshape B\kern-0.1em{\small A}\kern-0.1em B\kern-0.1em{\small A\kern-0.2em R}}}

\def\mass{4143.0\pm2.9(\mathrm{stat})\pm1.2(\mathrm{syst})~\MeVcc }
\def\width{11.7^{+8.3}_{-5.0}(\mathrm{stat})\pm3.7(\mathrm{syst})~\MeVcc }
\def\yield{14\pm5 }
 
\def\massdifffit{1046.3\pm2.9~\MeVcc }
\def\widthfit{11.7^{+8.3}_{-5.0}~\MeVcc }

      [1999/12/01 v1.4c Il Nuovo Cimento]
\documentclass{cimento}

\usepackage{graphicx}  
\title{Heavy Flavor Spectroscopy at the Tevatron }
\author{Kai Yi for the D0 and CDF collaborations\from{ins:x}\ETC
}
\instlist{\inst{ins:x} Department of Physics and Astronomy, University of Iowa, Iowa City, IA 52242, USA}
\PACSes{\PACSit{14.40.Nd, 14.40.Lb, 14.40.Pq}{} }
\begin{document}

\maketitle

\begin{abstract}
The Tevatron experiments have each accumulated about 6~\ifb ~of good data since the start of Run II. 
This large dataset provides excellent opportunities for heavy flavor spectroscopy studies 
at the Tevatron. This article will cover the latest $\Upsilon(nS)$ polarization 
studies as well as exotic meson spectroscopy results.

\end{abstract}

\section{Heavy Baryon--$\Omega_b$}
\indent

Here we discuss the most recent observation of $\Omega_b (bss)$ by both the D0 (1.3~\ifb ~of data)
and CDF (4.2~\ifb ~of data) experiments~\cite{omegab}. Both experiments observe this state through the 
following decay channel: $\Omega_b^-\rightarrow J/\psi\Omega^-; J/\psi\rightarrow
\mu^+\mu^-, \Omega^-\rightarrow \Lambda K^-; \Lambda\rightarrow p\pi$. 
Charge conjugate modes are included implicitly in this note.
D0 used a boosted decision tree to reconstruct the $\Omega$ signal, while CDF 
used the traditional cut-based selection to reconstruct the $\Omega$ signal. 
The reconstructed $\Omega_b$ mass plots 
from the two experiments are shown in Fig.~\ref{omegab}.
However, the $\Omega_b$ mass measured by D0 
($6165 \pm 10 (stat) \pm 13 (syst)$ MeV/$c^2$) and CDF ($6054.4 \pm 6.8 (stat) \pm 0.9 (syst)$ MeV/$c^2$) experiments 
disagree at the level of 6$\sigma$.
The measured relative branching fraction with $\Xi_b$ is also different 
at a level of $1.3\sigma$ between D0 ($0.80\pm0.32^{+0.4}_{-0.22}$) and 
CDF ($0.27\pm0.12\pm0.01$). D0 is working on an update with much more data 
to resolve the issue.

\begin{figure}[htb]
\centering
\includegraphics*[width=50mm]{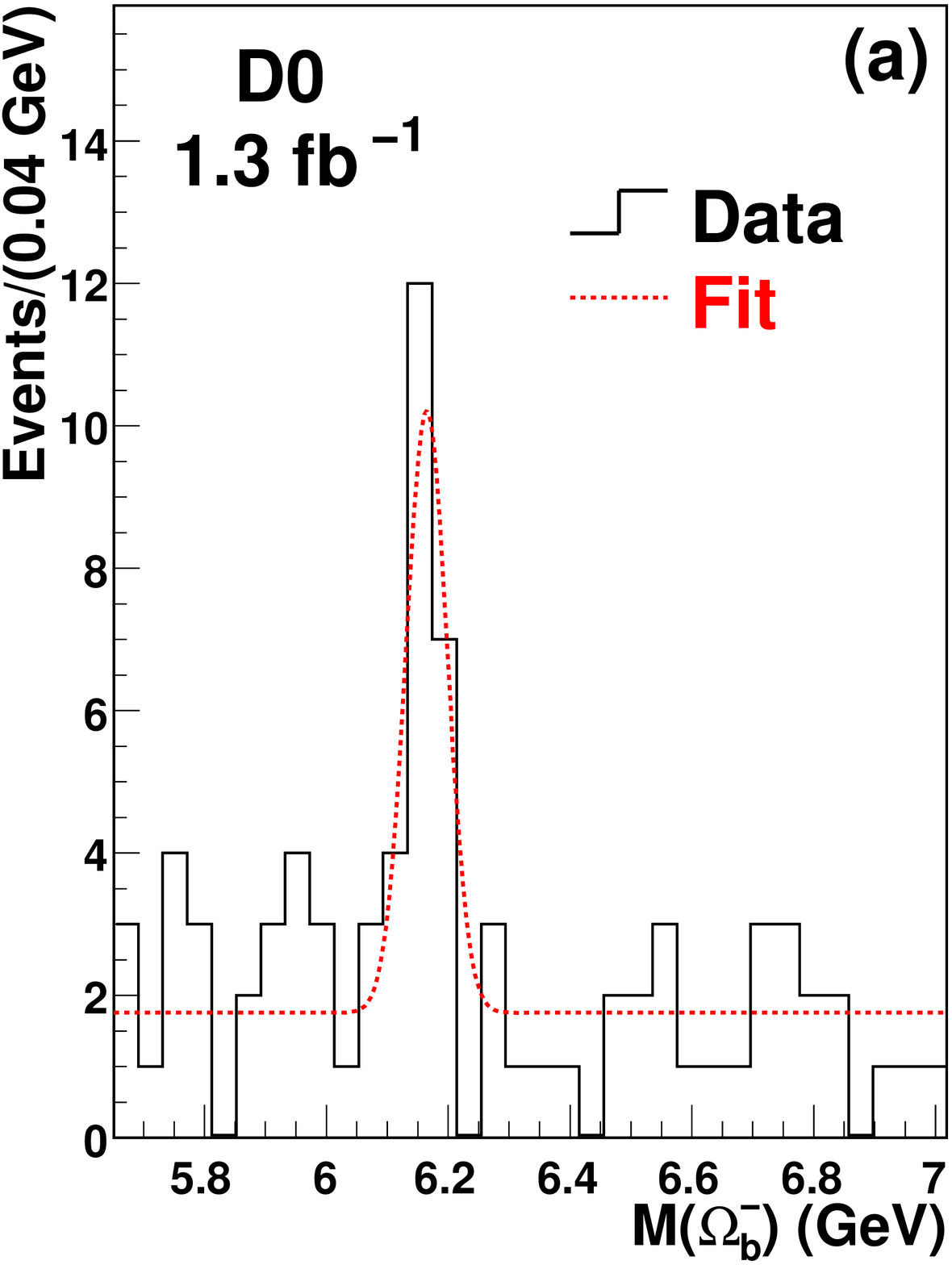}
\includegraphics*[width=68mm]{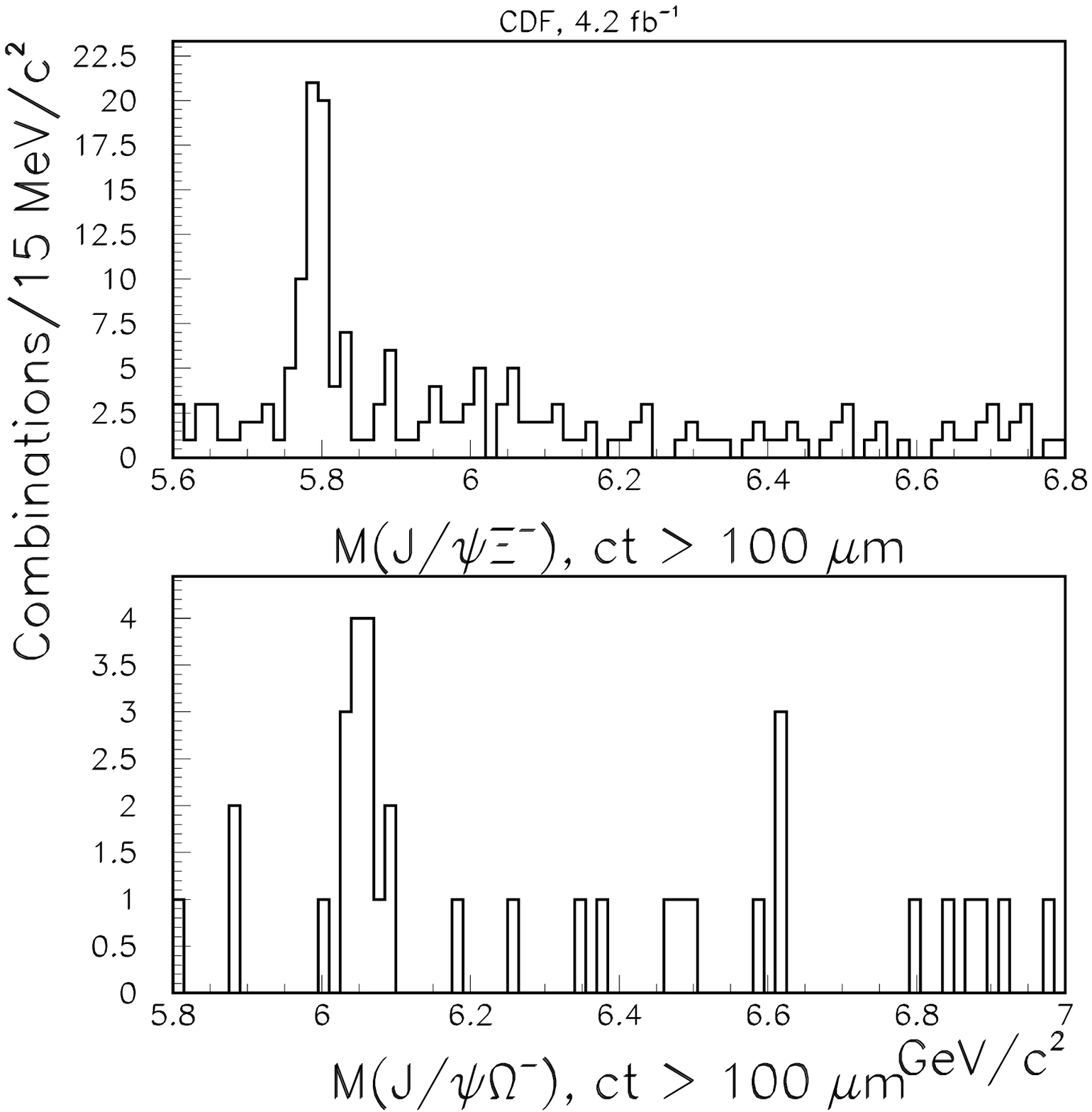} 
\caption{ The $\Omega_b$ mass spectrum from D0 (left), and CDF (right).
}
\label{omegab}
\end{figure}

\section{$\Upsilon(nS)$ polarization}
\indent

Vector meson production and polarization in hadronic collisions are usually 
discussed within the framework of
non-relativistic QCD (NRQCD). The theory predicts~\cite{vectortheory} that the 
vector meson polarization should become transverse in the perturbative 
regime; {\it i.e.}, at large 
transverse momentum $p_T$ of the vector meson. However, the  prediction is  
not supported by experimental 
observations~\cite{vectorpolarization}. We describe new results on this topic 
from the Tevatron. We define a parameter--$\alpha$ to measure the polarization:
\begin{equation}
\frac{d\Gamma}{dcos\theta^*}\propto  1+\alpha cos^2\theta^*
\end{equation}
where $\theta^*$ is the $\mu^+$ angle with respect to the $\Upsilon(nS)$ direction in the 
lab frame.
If the  meson is fully polarized in the transverse direction, $\alpha$ = 1. 
If it is fully aligned longitudinally, $\alpha$ =-1.

Fig.~\ref{Yns1} shows the comparison between the theoretical prediction 
of $\Upsilon(1S)$ (colored band) and the new CDF (left)~\cite{cdfy1s} and 
D0 (right)~\cite{d0yns} experimental results. 
In the low $p_T$ region, CDF shows nearly-unpolarized events, which is  consistent 
with the
CDF Run I result~\cite{cdfy1sruni}; D0 shows partially  longitudinally polarized events.
At higher $p_T$ , the CDF results tend toward longitudinal polarization while the D0 
result indicates  transverse polarization. Both CDF and D0 results at high $p_T$ deviate 
from theoretical predictions. It will be interesting to investigate with more data 
and in some detail; {\it e.g.} study $\eta$ dependence since the CDF and D0 analyses have different 
$\eta$ acceptance.

\begin{figure}[htb]
\centering
\includegraphics*[width=60mm]{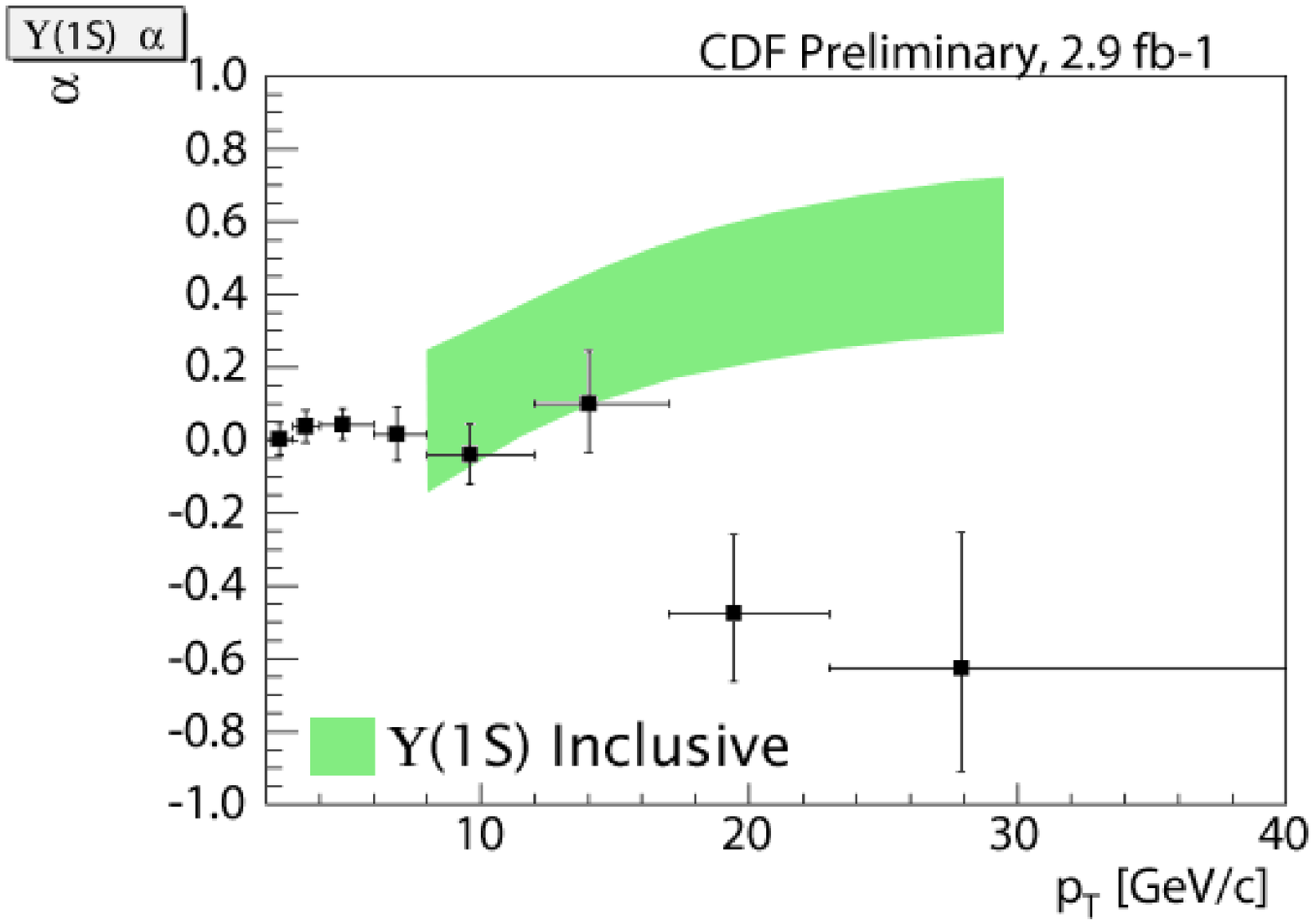}
\includegraphics*[width=60mm]{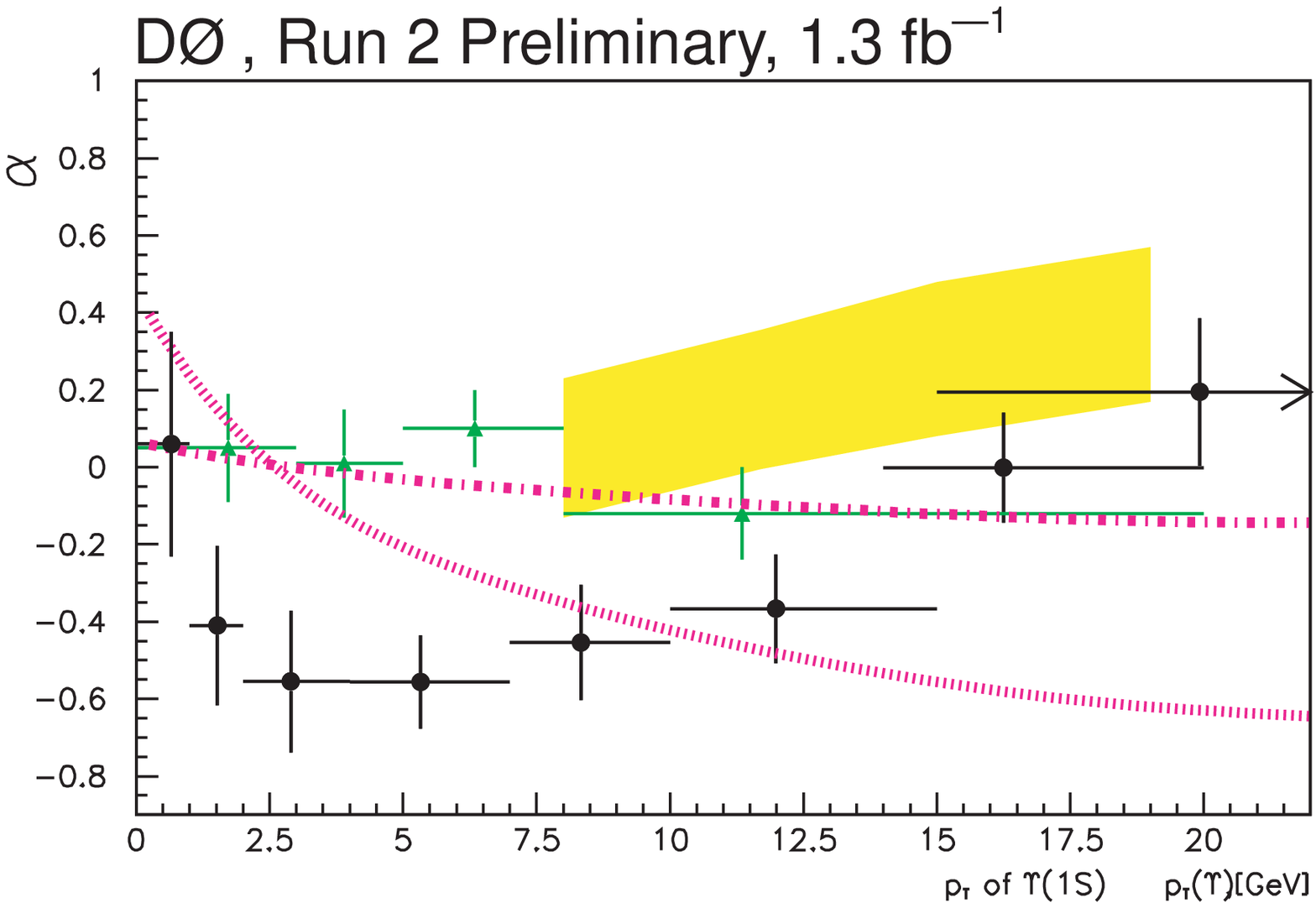}
\caption{ The polarization parameter $\alpha$ of $\Upsilon(1S)$ measured by 
CDF (left) and D0 (right, CDF I results are shown as green points).
}
\label{Yns1}
\end{figure}

\section{Exotic mesons}
\indent

It has been six years since the discovery of the $X(3872)$~\cite{x3872discovery}; 
however, the nature of this state 
has not yet been clearly understood. 
Due to the proximity of the $X(3872)$ to the $D^0D^{*0}$  threshold, 
the $X(3872)$ has been proposed as a 
molecule composed of $D^0$ and $D^{*0}$ mesons. The $X(3872)$ has also been speculated 
to be two nearby states, as in models such as the $diquark-antidiquark$ model. 
It is critical to make precise measurements of the mass and width of 
$X(3872)$ to understand its nature.
The  large  $X(3872)\rightarrow J/\psi\pi^+\pi^-$ sample accumulated at CDF enables a test of
the hypothesis that the $X(3872)$ is composed of two states and to make a 
 precise mass measurement of $X(3872)$ if it is consistent with a one-state hypothesis.

There are many more states, similar to $X(3872)$, that have
charmonium-like decay modes 
but are difficult to place in the overall charmonium 
system~\cite{y3940,y4260discovery,y4320}.  These unexpected new 
states have introduced challenges to the conventional $q\bar{q}$ meson model 
and revitalized interest in exotic mesons in the charm 
sector~\cite{conventional}, although the existence of exotic mesons
 has been discussed for many years~\cite{PDG}. 
Until recently all
of these new states involved 
only $c$ quark  and light quark ($u$, $d$) decay products.   
The $J/\psi\phi$ final state enables us to extend the exotic meson searches  
to $c$ quark and heavy $s$ quark decay products. 
An investigation of the $J/\psi\phi$ system produced in exclusive 
$B^+\rightarrow J/\psi\phi K^+$ decays with $J/\psi \rightarrow \mu^+ \mu^-$ 
and $\phi\rightarrow K^+K^-$ is reported here.

\subsection{Measurement of the mass of $X(3872)$}
\indent

A CDF analysis tested the hypothesis 
of whether the observed X(3872) signal is composed of  two different states 
as predicted in some four-quark models  using the CDF inclusive $X(3872)$ sample. 
The  $X(3872)$ mass signal is fit with a Breit-Wigner function convoluted with a 
resolution function~\cite{x3872mass}. 
Both functions contain a width scale factor that is a free parameter in the fit 
and therefore sensitive to the shape of the mass signal. The measured width scale factor is 
compared to the values seen in simulations which assume two states with the given mass 
difference and ratio of events. The resolution in the simulated events is corrected for the 
difference between data and simulation as measured from the $\psi(2S)$.
The result of this hypotheses test shows that the data is  consistent with  a single 
state. Under the assumption of two states with equal amount of observed events, a limit of
$\Delta m < 3.2 (3.6)$ \MeVcc~ is set at 90\% (95\%) C.L.

Since the $X(3872)$ is consistent with one peak in our test, 
its mass is measured in an 
unbinned maximum likelihood fit. The systematic uncertainties are determined from the difference 
between the measured $\psi(2S)$ mass and its world average value, the potential variation of the 
$\psi(2S)$ mass as a function of kinematic variables, and the difference in Q value 
between $X(3872)$  and $\psi(2S)$. Systematics 
due to the fit model are negligible. The measured $X(3872)$ mass is:
$m(X(3872)) = 3871.61 \pm 0.16 (stat) \pm 0.19 (syst)$ \MeVcc, which  
is the most precise measurement to date, as shown in Fig.~\ref{Fig8}~\cite{x3872mass,x3872massbelle}.

\begin{figure}[htb]
\centering
\includegraphics*[width=70mm]{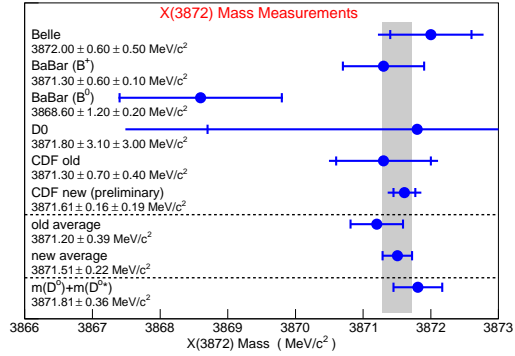}
\caption{
An overview of the measured $X(3872)$ masses from the experiments observing the $X(3872)$.
}
\label{Fig8}
\end{figure}

\subsection{Evidence for Y(4140) }
\indent

\begin{figure}[htb]
\centering
\includegraphics*[width=70mm]{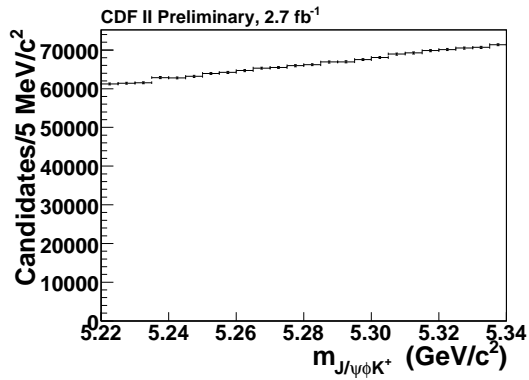}
\caption{The  $J/\psi\phi K^+$ mass before   
minimum ${L_{xy}(B^+)}$ and  kaon $LLR$ requirements.
}
\label{Fig1}
\end{figure}

The procedure for this analysis is to reconstruct the $B^+\rightarrow J/\psi\phi K^+$ signal and 
then search for structures in the $J/\psi\phi$ mass spectrum~\cite{y4140}.
The $J/\psi\rightarrow \mu^+\mu^-$ events are recorded using a dedicated dimuon trigger. 
The  $B^+\rightarrow J/\psi\phi K^+$ candidates are reconstructed 
by combining a $J/\psi\to\mu^+\mu^-$ candidate, a $\phi\rightarrow K^+K^-$ candidate,  
and an additional charged track. 
Each track  is required to have at least 4  axial silicon  hits and have a transverse momentum 
greater than 400 \MeVc. 
The reconstructed mass of each vector meson candidate must lie within 
a suitable range from the nominal 
values ($\pm$50 \MeVcc~ for the $J/\psi$ and $\pm$7 \MeVcc~ for the $\phi$).  In the final $B^+$ 
reconstruction the $J/\psi$ is mass constrained, and the $B^+$ candidates must have
$p_T > 4$ \GeVc.
The $P(\chi^2)$ of the mass- and vertex-constrained fit to the 
$B^+\rightarrow J/\psi\phi K^+$ candidate is required to be greater than 1\%.

\begin{figure}[htb]
\centering
\includegraphics*[width=70mm]{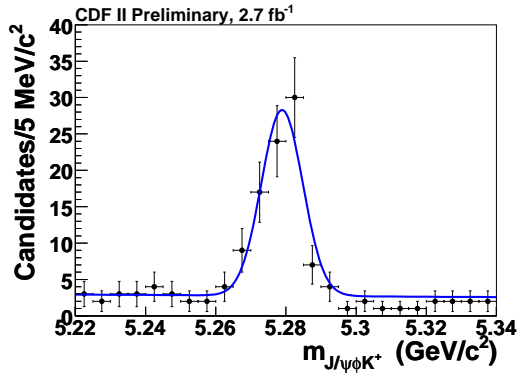}
\caption{
The $J/\psi\phi K^+$ mass after  
minimum ${L_{xy}(B^+)}$ and $LLR$ requirements; the solid line is a fit to 
the data with a Gaussian signal 
function and linear background function.
}
\label{Fig2}
\end{figure}

To suppress combinatorial background, ${dE/dx}$ and Time-of-Flight
($\mathrm{TOF}$) information is used to identify all three kaons in the final state.
The information is summarized in a log-likelihood ratio ($LLR$), which 
reflects how well a candidate track can be positively identified 
as a kaon relative to other hadrons.
In addition, a minimum ${L_{xy}(B^+)}$ is required for the 
$B^+\rightarrow J/\psi\phi K^+$ candidate,
where ${L_{xy}(B^+)}$ is the projection onto ${\vec{p}_T(B^+)}$ 
of the vector connecting the primary vertex to the ${B^+}$ decay vertex.   
The  ${L_{xy}(B^+)}$ and   ${LLR}$ requirements for $B^+\rightarrow J/\psi\phi K^+$ 
are then chosen to 
maximize $\mathit{S/\sqrt{S+B}}$,  
where $\mathit{S}$ is the number of $B^+\rightarrow J/\psi\phi K^+$ signal events 
and $\mathit{B}$ is the number of background events  implied from the 
$B^+$ sideband.  
The requirements obtained by  maximizing $\mathit{S/\sqrt{S+B}}$ are ${L_{xy}(B^+)}>500 ~\um$ 
and  ${LLR}>0.2$.

\begin{figure}[htb]
\centering
\includegraphics*[width=70mm]{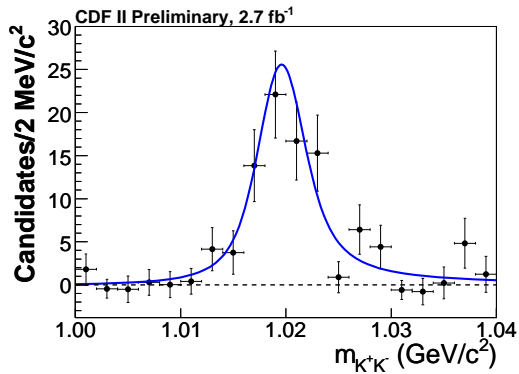}
\caption{
The $B^+$ sideband-subtracted $K^+K^-$ mass without the $\phi$ mass window requirement.
The solid curve is a $P$-wave relativistic Breit-Wigner fit to the data. 
}
\label{Fig3}
\end{figure}

The invariant mass of $J/\psi\phi K^+$, after $J/\psi$ and $\phi$ mass window  
requirements, before and after the minimum ${L_{xy}(B^+)}$ and kaon ${LLR}$ requirements, 
are shown in Fig.~\ref{Fig1} and  Fig.~\ref{Fig2}, respectively. 
The $B^+$ signal is not distinguishable 
before the ${L_{xy}(B^+)}$ and kaon ${LLR}$ requirements are applied, but a clear 
$B^+$ signal is seen after the  requirements. 
A fit with a Gaussian signal function and a 
linear background function  to the mass spectrum of $J/\psi\phi K^+$ (Fig.~\ref{Fig2}) 
returns a $B^+$ signal of $75\pm10(\mathrm{stat})$ events.
The ${L_{xy}(B^+)}$ and ${LLR}$  requirements reduce the background by a factor of 
approximately 20 000 
while keeping a signal efficiency of approximately 20\%.
The $B^+$ signal candidates are selected with a mass 
within  3$\sigma$  of the nominal $B^+$ mass; 
the purity of the $B^+$ signal in that mass window is about 80\%.

\begin{figure}[htb]
\centering
\includegraphics*[width=70mm]{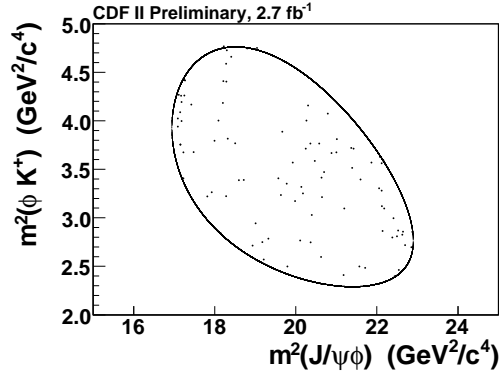}
\caption{
The Dalitz plot of ${m^2(\phi K^+)}$ 
versus ${m^2(J/\psi \phi)}$  in the  $B^+$ 
mass window. The boundary shows the kinematically-allowed region.
}
\label{Fig4}
\end{figure}

The combinatorial background under the $B^+$ peak includes
$B$ hadron decays such as $B^0_s \rightarrow \psi(2S)\phi \rightarrow J/\psi \pi^+\pi^-\phi$,
in which the pions are misidentified as kaons.  However,
background events with misidentified kaons cannot
yield a Gaussian peak at the $B^+$ mass consistent with the
5.9 \MeVcc mass resolution.  
Figure~\ref{Fig3} shows the  $K^+ K^-$ mass  
from $\mu^+\mu^-K^+K^- K^+$ candidates within $\pm 3 \sigma$ of the nominal $B^+$ mass
with $B$ sidebands subtracted  before applying the $\phi$ mass window requirement.
Using a smeared $P$-wave relativistic Breit-Wigner (BW)~\cite{pbw} line-shape fit to the spectrum 
returns a $\chi^2$ probability of 28\%.  
This shows that 
the $B^+ \rightarrow J/\psi K^+K^-K^+$ final state is well described 
by $J/\psi \phi K^+$.

\begin{figure}[htb]
\centering
\includegraphics*[width=70mm]{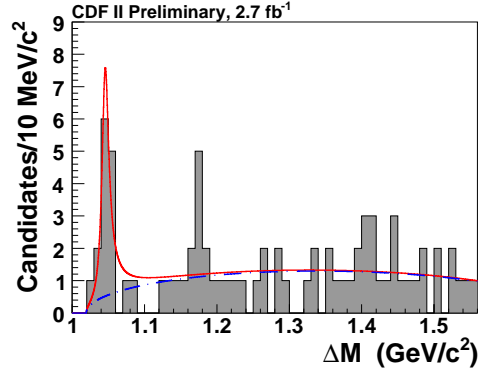}
\caption{
 The mass difference, $\Delta M$, between $\mu^+\mu^-K^+K^-$ and $\mu^+\mu^-$,  
 in the $B^+$ mass window. 
The dash-dotted curve is 
the background contribution and the red solid curve is the total unbinned fit.
}
\label{Fig5}
\end{figure}

The effects of detector acceptance and selection
  requirements are examined using 
$B^+ \rightarrow J/\psi \phi K^+$ MC events simulated  by a phase-space distribution. 
The MC events are smoothly distributed in the Dalitz plot and 
in the $J/\psi \phi$ mass spectrum. No artifacts were observed from MC events. 
Figure~\ref{Fig4} shows the Dalitz plot of ${m^2(\phi K^+)}$ 
versus ${m^2(J/\psi \phi)}$,   
and  Fig.~\ref{Fig5} shows 
the mass difference, $\Delta M= m(\mu^+\mu^-K^+K^-)- m(\mu^+\mu^-)$, for events in 
the $B^+$ mass window in our data sample. 
The enhancement in the $\Delta M$ spectrum just above $J/\psi\phi$ threshold is examined. 
The high--mass part of the spectrum beyond $1.56$ \GeVcc~  is excluded to avoid 
combinatorial backgrounds that would be expected from misidentified 
$B^0_s\rightarrow \psi(2S) \phi\rightarrow (J/\psi\pi^+\pi^-)\phi$ decays.
The enhancement is modeled by an $S$-wave relativistic BW 
function~\cite{sbw} convoluted with 
a Gaussian resolution function with the RMS fixed to 1.7 \MeVcc~obtained from MC,  
and three--body phase space~\cite{PDG}  is used to describe the background shape.
An unbinned likelihood fit to the $\Delta M$ distribution,   
as shown in Fig.~\ref{Fig5}, returns a yield of $\yield$ events,   
a $\Delta M$ of $\massdifffit$, and a width of $\widthfit$.
To investigate possible reflections, the Dalitz plot and 
projections into  the $\phi K^+$ and $J/\psi K^+$ spectra are examined.  
No evidence for any other structure in the $\phi K^+$ and $J/\psi K^+$ 
spectra is found.

The log-likelihood ratio of $-2{\mathrm{ln}}(\mathcal{L}_0/\mathcal{L}_{{max}})$~
is used to determine the significance of the enhancement, 
where $\mathcal{L}_0$ and $\mathcal{L}_{{max}}$ are the likelihood 
values for the null  hypothesis fit and signal hypothesis fit.
The $\sqrt{-2{\mathrm{ln}}(\mathcal{L}_0/\mathcal{L}_{{max}})}$ value 
is 5.3 for a pure three--body phase space 
background shape assumption. 
  Using the background distribution alone,  $\Delta M$ spectra are generated, 
  and searched for the most significant fluctuation with 
$\sqrt{-2{\mathrm{ln}}(\mathcal{L}_0/\mathcal{L}_{{max}})}\ge 5.3$  in
  each spectrum in the mass range of 1.02 to 1.56 \GeVcc, with widths
  in the range of 1.7 (detector resolution) to 120 \MeVcc~(ten times the observed width).

The resulting $p$-value  from 3.1 million simulations 
is $9.3\times 10^{-6}$, corresponding to a significance of 4.3$\sigma$.
This process is repeated with a flat combinatorial non-B background and three--body PS 
for non-resonance $B$ background, which gives a significance of 3.8$\sigma$.

One's eye tends to be drawn to a second cluster of events
around 1.18 \GeVcc~in Fig.~\ref{Fig5}. 
This cluster is  close to one pion mass
above the peak at the $J/\psi\phi$ threshold. However, this cluster 
is statistically insufficient to infer the presence of a second structure.

\section{Summary}
  \indent

Both D0 and CDF observed the $\Omega_b$ baryon through the same reconstruction channel.
However, the measured $\Omega$ mass disagrees at a level of 6$\sigma$ between 
the two experiments. D0 is working on an update with much more data 
to resolve this issue.

For $\Upsilon(1S)$ polarization, 
CDF result shows nearly-unpolarized events at low $p_T$, 
while D0 shows partially  longitudinally polarization.
At higher $p_T$, CDF results tend toward longitudinal polarization while D0 results indicate 
transverse polarization. Both CDF and D0 results at high $p_T$ deviate 
from theoretical predictions. 
CDF is continuing the analysis and will double the
dataset.  D0 has the opportunity to study the rapidity dependence, since
their measurement spans the range $|y|<1.8$ compared to 0.6 for CDF.

Studies using CDF's $X(3872)$ sample, the largest in the world,
indicate that the $X(3872)$ is consistent with the 
one-state hypothesis and this leads to the most precise mass measurement of $(X3872)$. 
The value is below, but within the uncertainties of 
the $D^{*0}D^0$ threshold. The explanation 
of the X(3872) as a bound D*D system is therefore still an option.

The $B^+ \rightarrow J/\psi \phi K^+$   sample at CDF enables a
 search for structure in the $J/\psi\phi$ mass spectrum,  
and evidence is found for a narrow structure near the $J/\psi\phi$ 
threshold with a significance estimated to be at least  3.8$\sigma$.
Assuming an $S$-wave relativistic BW, the mass (adding $J/\psi$ mass) and width of this structure, 
including systematic  uncertainties,  are measured to be $\mass$ 
and $\width$, respectively.   This structure does not 
fit conventional expectations for a charmonium state 
because as a $c\bar{c}$ state it is expected to have a tiny branching ratio 
to $J/\psi\phi$ with its  mass well beyond open charm pairs.
The new structure is termed the $Y(4140)$. The branching ratio  
of $B^+\rightarrow Y(4140) K^+, Y(4140)\rightarrow 
J/\psi\phi $ is estimated to be $9.0\pm3.4(stat)\pm2.9(B_{BF}))\times10^{-6}$.

\acknowledgments

We thank the Fermilab staffs and the technical staffs of the participating institutions 
for their vital contributions.

\end{document}
\endinput